# Urban Green Index estimation based on data collected by remote sensing for Romanian cities

**Marian NECULA**[1] (neculamarian18@stud.ase.ro)
Economic Cybernetics and Statistics Doctoral School,
The Bucharest University of Economic Studies

**Tudorel ANDREI, Ph.D.** (andreitudorel@csie.ase.ro)
Faculty of Cybernetics, Statistics and Economic Informatics,
The Bucharest University of Economic Studies

**Bogdan OANCEA, Ph.D.** (bogdan.oancea@faa.unibuc.ro)
The Faculty of Business and Administration, University of Bucharest

**Mihaela PĂUN, Ph. D.** (mihaela.paun@faa.unibuc.ro, mihaela.paun@incdsb.ro)
The Faculty of Business and Administration, University of Bucharest /
Bioinformatics Department, National Institute for R &D for Biological Sciences

## ABSTRACT

The modernization of official statistics involves the use of new data sources, such as data collected through remote sensing. The document contains a description of how an urban green index, derived from the SDG 11.7 objective, was obtained for Romania's 41 county seat cities based on free data sets collected by remote sensing from the European and North American space agencies. The main result is represented by an estimate of the areas of surfaces covered with vegetation for the 40 county seat towns and the municipality of Bucharest, relative to the total surface. To estimate the area covered with vegetation, we used two data sets obtained by remote sensing, namely data provided by the MODIS mission, the TERRA satellite, and data provided by the Sentinel 2 mission from the Copernicus space program. Based on the results obtained, namely the surface area covered with vegetation, estimated in square kilometers, and the percentage of the total surface area or urban green index, we have created a national top of the county seat cities.

---

1. corresponding author.

# 1. INTRODUCTION

Land surface monitoring based on satellite imagery has been used since the 1960s and 1970s and has been a prolific source of important data for an extremely wide range of users. With initial applications in the military and meteorological fields, the monitoring capabilities have been widely expanded, for example the Sentinel missions within the Copernicus program (ESA, 2022a) generated in 2021 approximately 32 petabytes of data. In this context, there is an increased interest from the official statistics to explore and possibly incorporate this data source into the statistical production. Thus, within the ESSnet on Big Data II project, a pilot was carried out to explore the potential of remote sensing data (Work Package H, 2021) supported by several case studies, among which we list applications regarding the monitoring of surfaces, agricultural crops and natural vegetation, applications regarding the monitoring of environmental indicators included in sustainable development strategies, etc. Also, the the United Nations Statistics Division, through projects exploring Big Data data sources (Task Team on Earth Observations, 2017), estimates that this data source has a growing potential to be used in statistical production, either in complementing existing statistics or developing innovative statistics in response to the increased demand from data users.

Another argument for researching the potential of data sources collected through remote sensing derives from the modernization strategy of the National Institute of Statistics (INS, 2022) which considers the diversification of data sources and methods. The main limitation identified in the development of such projects is represented by the human resource that must have knowledge both in the field of official statistics and in the computational methods related to the exploitation of the new data sources.

Within the Sustainable Development Goal 11.7 regarding universal access to green public spaces, indicator 70 was proposed regarding the proportion of green spaces relative to the total area of urban settlements, in order to monitor the quality of life within human settlements classified as cities and to ensure equitable access to green infrastructure. The implementing of such an indicator is built on the assumption that a city whose green infrastructure is maintained and increased in relation to the total area, is translated into economic and social effort to ensure resilience to extreme climatic phenomena and improve the quality of life, e.g. reducing the effects of pollution, reducing crime, increasing residents' satisfaction, etc. (UN, 2022). At the European level, there are many initiatives regarding the sustainable development of cities and the conservation/expansion of green infrastructure and ecosystems, among which we mention the Urban Agenda (European Commission, 2016), the European Strategy on Green Infrastructure (European Commission, 2013) or the decision of the European Parliament on the General Union Environment Action Program to 2020 (European Parliament 2013). Among the European initiatives for monitoring urban surfaces, the most important, from our point of view, is represented by the Land Surface Monitoring Service, within the Copernicus program of the European Commission developed in partnership with the European Space Agency (European Commission, 2022). The Copernicus program provides users, free of charge, with access to data obtained through remote sensing, through the Sentinel space missions, but also to data obtained through in situ collection (ESA, 2022c). Green urban spaces, according to the definition presented in the Mapping and Assessment of Ecosystems and their Services report of the European Commission (European Commission 2014), represent land surfaces fully or partially covered with vegetation, components of the green infrastructure of cities. The definition used does not specify what type of vegetation covers the respective surfaces and does not functionally address the type of use of the green space, i.e. whether it is public or private property, whether it is vegetation of natural origin or as a result of human development, whether the type of vegetation is interspersed between buildings and infrastructure or is continuous in nature, etc. Thus, two types of classifications of urban green spaces are presented that can contain the following atomic elements:

- green buildings (e.g. the roof is covered with vegetation);
- private property (alleys, singular trees, etc.);
- public gardens and parks;
- lands covered with vegetation intended for agriculture;
- lands covered with natural vegetation;
- areas of the region between water and land (shore, coastal areas, etc.).

Within the classification used in the Copernicus program (ESA, 2022b), of data collected in-situ through the CORINE Land Use Land Cover statistical research, green urban spaces are named green urban areas and can include: parks; ornamental gardens; private properties landscaped with vegetation; botanical and zoological gardens; public squares covered with vegetation; green spaces between glades; cemeteries; areas with vegetation for recreational purposes; etc.

This classification does not include: agricultural land included in the urban area, cemeteries outside the urban area and other possible types of surfaces covered with vegetation.

## 2. DATA AND METHODS

The present study uses two sources of remote sensing data, namely:

• Hyperspectral data Terra MODIS (Moderate resolution imaging spectroradiometer) were accessed through the Earthdata service (NASA, 2022a), an archive for accessing and distributing a large collection of remote sensing data. There are other web services through which MODIS data can be accessed and downloaded. The data are provided in the hierarchical data format – hdf5- (HDF Group, 2022), a scientific data transmission standard independent of the hardware or software architecture of computing machines. The MODIS hdf5 file contains the NDVI (Normalized Difference Vegetation Index) data and associated spatial attributes. The file contains the NDVI as well as the Enhanced Vegetation Index (EVI). More details about the NDVI vegetation index are presented later in this chapter. On average, the size of a MODIS hdf5 file for NDVI is between 150 and 250 megabytes (covering an area of about 5500 km^2). The period for which we downloaded the data is July 2022.

• Sentinel-2 multispectral data were accessed through the Open Access Hub (ESA, 2022d). The data is provided in SENTINEL-SAFE format (ESA, 2022e), a .zip archive containing metadata files (requirement of data preprocessing, data quality, etc.) and the actual image files in .jp2 (JPEG2000) format for those 13 spectral bands and other pre-calculated products (cloud mask, pixel value quality mask, surface classification mask, etc.). On average, the size of such an archive is about 1 gigabyte (referred to an area of 100km^2). The period for which the data was downloaded is between 01-Jun-2022 and 31-Jul-2022.

MODIS is one of the instruments for measuring the electromagnetic radiation reflected by the earth's surface, installed on the Terra satellite (NASA, 2022b) launched and operated by NASA in 1999. The objective of the mission is to monitor the earth's surface and atmosphere. The Terra satellite covers the entire Earth's surface on average every two days and through MODIS provides images at a spatial resolution of 250, 500 and 1000 meters. The instruments are capable of recording electromagnetic radiation (36 spectral bands) between 400 and 14400 nm. The data provided is pre-processed and prepared for use in scientific or other analyzes with minimal effort. MODIS datasets include, among others: vegetation indices; land surface temperature and temperature anomaly detection (fires); the reflectance of the earth's surface. Certain data sets are available daily, depending on the degree of pre-processing or spatial/temporal coverage required by users.

Sentinel-2 is one of the space missions of the European Space Agency, with the objective of monitoring land surfaces (ESA, 2022a). The mission consists of two satellites launched between 2016 and 2017, Sentinel-2A and Sentinel-2B, with polar orbits, phase-shifted by 180 $[\char`\^\circ]$, which have an average revisit time at the Equator of about 5 days. The satellites are equipped with a MultiSpectral Instrument (MSI), which can detect electromagnetic radiation between 400 and 2200 nm (13 spectral bands). The instrument can provide data at a spatial resolution between 10m and 60m. In Romania, the duration between two consecutive visits to the same area is about 10 days, on average, or about 3-4 times a month. The big disadvantage of the two instruments, in particular, the one installed on the Sentinel-2 satellites, is represented by the surfaces masked by clouds or extreme atmospheric effects. For both data sources, and/or other similar data sources, data sets are available, according to a convention, on pre-processing levels necessary to perform a certain type of analysis (Level 1A, Level 1B, Level 1C, Level 2A, Level 2B, etc.). In this case we have retrieved L2A (Level 2A) data sets, which have been corrected by the data provider, both geometrically and optically, and can be used directly in the analyses. Thus, the object of this analysis is represented by a dimensionless physical quantity called spectral (directional) reflectance, which represents the ratio between the radiance reflected by the surface of a material and the radiance incident on that surface and directly depends on the type/material of the surface (ISO, 1989).

In figure 1, the image obtained by the MSI Sentinel 2 instrument is represented graphically, by merging the spectral bands related to the reflectance in the visible range. When passing through the Sentinel 2 observation areas, it covers an area with an average width of 290 km.

**Sentinel 2 MSI related spectral bands in the visible range (RGB)**

*Figure 1*

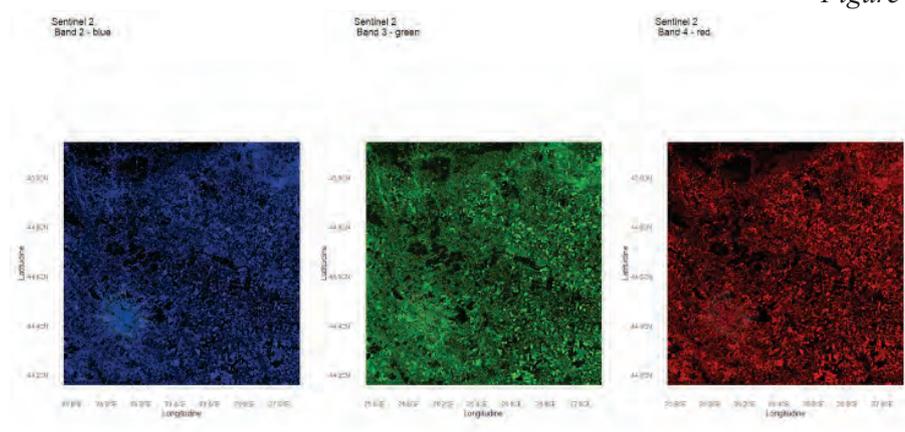

**Composite RGB image for Sentinel 2  
(June 2022, area T35TMK - Bucharest)**

*Figure 2*

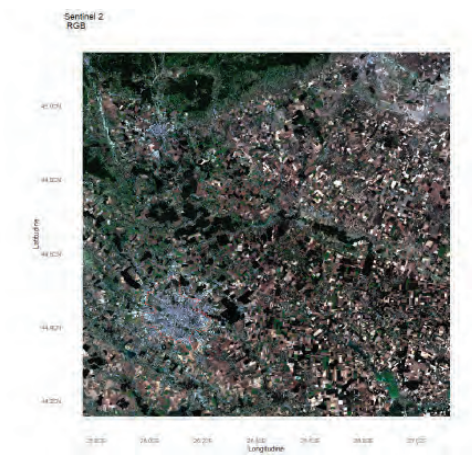

In the case of MODIS data, we downloaded the dataset containing the green index, represented in figure 3 for the entire surface of the country. The image was obtained by mosaic between area 194 and 204 related to the cartographic design mode used by MODIS. As a side note, MODIS covers an area 2330 km wide in a single pass.

**MODIS image - normalized difference vegetation index  
(June 2022, area 194 + area 204 Romania)**

*Figure 3*

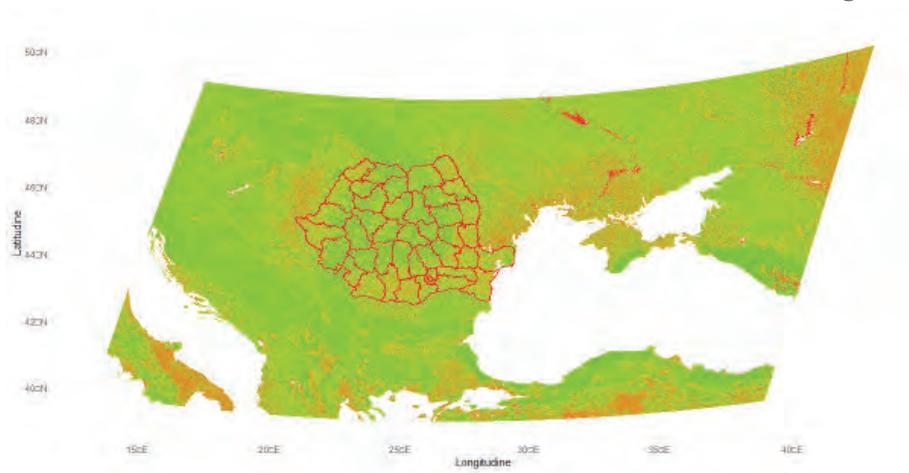

As auxiliary data, we used the cartographic delimitation of the county-seat municipalities, for all 41 counties and the municipality of Bucharest, graphically represented in figure 4. The delimitation used does not include the territorial administrative unit with the same name, and refers strictly to the urban, specific urban area component city of the county seat. For example, the territorial administrative unit Alba Iulia is composed of the city of Alba Iulia (within and outside the built-up area), Bărăbanț, Micești, Oarda și Pâclișa, from which we strictly choose the city of Alba Iulia.

The total data size is approximately 80 Gb for Sentinel 2 and 300 Mb for Modis. In the case of Sentinel 2, it was necessary to increase the temporal period of interest to 2 months, in order to identify data sets that do not contain the missing date or whose quality is not affected by the extreme presence of cloud cover of the surfaces of interest.

**Romania. County residences and the municipality of Bucharest.**

*Figure 4*

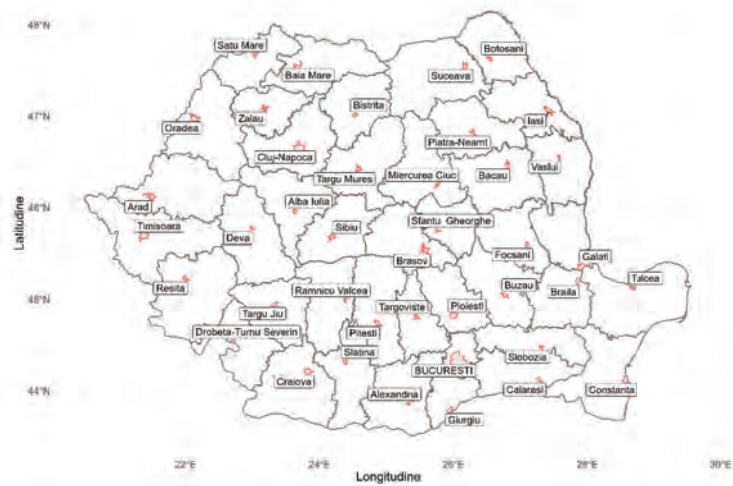

The normalized difference vegetation index is a popular measure in remote sensing data analysis applications for the identification/visualization of vegetated areas (Weier and Herring, 2000). The index is a dimensionless quantity, with values in the closed interval [-1, 1], quantifying the degree of vegetation coverage of the earth's surface. According to Huete, Justice and van Leeuwen (1999), the normalized difference vegetation index was created based on empirical observations of the interaction between electromagnetic waves, light in the red visible spectrum and near infrared spectrum, and vegetation.

Through the measurements, an increase in the intensity of radiation in the near infrared zone and the absorption of red light from the visible spectrum was observed. NDVI has some ability to discriminate between the types of vegetation covering an area (agricultural vegetation, forest, shrubs, etc.), as well as the quality of that vegetation (dry, green, etc.). The sensor-independent formula is:

$$NDVI = \frac{NIR - RED}{NIR + RED}$$

where
*NIR = the associated near-infrared wavelength between 750 and 1400 nm*
*RED = the wavelength associated with red visible range between 620 and 750 nm.*

For MODIS, the spectral bands associated with the two wavelengths are band 1, red visible range, and band 2, near infrared, and in the case of Sentinel 2, band 4, red visible range, and band 8, near infrared. However, there are differences between the calculation methods.
MODIS NDVI is already calculated based on the following algorithm:
*Step 1: The NDVI index is estimated for each MODIS sensor record over an area of interest.*
*Step 2: The composite NDVI index for 16 days at a resolution of 250m per pixel is selected as the maximum of the NDVI value associated with the same pixel (spatial coordinates), and of course, the value is not affected by measurement errors, either due to the sensor or external factors.*

The MODIS NDVI composite index is used in the paper. In the case of Sentinel 2, the index is calculated for a single record (calendar date) based on the previously described calculation formula and the availability of an image that is not affected by measurement errors, or excessive presence of clouds.

**Some NDVI intervals for vegetation discrimination**

*Table 1*

| Category | Vegetation sub-category | STATCAN (Lantz, Grenier, Wang, 2021) | Weier and Herring, 2000 (sensor NOAA AVHRR) | ESA 2022f |
|---|---|---|---|---|
| Non-vegetation | N/A | [-1, 0.5] | [-1, 0] | <= 0.4 |
| Vegetation | - | [0.5, 1] | (0, 1] | >0.4 |

In order to select an optimal threshold for the discrimination between areas covered with vegetation and those with non-vegetation, we performed manual comparisons between the images represented in natural (visible) colors provided by the Google Maps service (satellite image layer) and different intermediate discrimination thresholds starting from literature data, respectively the 0.5 to 0.7 NDVI threshold for MODIS, and the 0.3 to 0.6 NDVI threshold for Sentinel-2, both thresholds built with a step of 0.05. In table 1. The values for the discrimination threshold for each county seat city and the average used for the final calculation of the areas covered with vegetation are presented. Thus, for MODIS we obtained a value close to 0.58 and for Sentinel 2 we obtained 0.4.

From figure 5, we can see the sensitivity of the results to the selection of the discrimination threshold between the surfaces, the threshold being in an inversely proportional, almost linear relationship with the vegetation area detected by the green index.

Figure 6 shows the histogram of NDVI values for the Bucharest. From the graph it can be seen that the distribution of NDVI values is similar, but, on the one hand, there are differences derived from the different resolution of the two sensors and, on the other hand, differences between the parameters of the two distributions, e.g. MODIS values are concentrated around the 0.5 point, and Sentinel 2 values tend, roughly, to a point in the range [0.3-0.4]. These values are similar to the values resulting from the manual determination of the discrimination threshold used to estimate the green index.

**Vegetation discrimination threshold – manually estimated**

*Table 2*

| Residence | Modis | Sentinel 2 |
|---|---|---|
| Alba Iulia | 0.6 | 0.35 |
| Alexandria | 0.55 | 0.4 |
| Arad | 0.6 | 0.4 |
| Bacau | 0.65 | 0.45 |
| Baia Mare | 0.65 | 0.4 |
| Bistrita | 0.6 | 0.4 |
| Botosani | 0.5 | 0.35 |
| Braila | 0.6 | 0.45 |
| Brasov | 0.65 | 0.45 |
| Bucuresti | 0.7 | 0.45 |
| Buzau | 0.55 | 0.35 |
| Calarasi | 0.55 | 0.35 |
| Cluj-Napoca | 0.65 | 0.4 |
| Constanta | 0.55 | 0.4 |
| Craiova | 0.6 | 0.35 |
| Deva | 0.65 | 0.45 |
| Drobeta-Turnu Severin | 0.5 | 0.4 |
| Focșani | 0.55 | 0.4 |
| Galati | 0.6 | 0.35 |
| Giurgiu | 0.55 | 0.4 |
| Iasi | 0.6 | 0.4 |
| Miercurea Ciuc | 0.55 | 0.35 |
| Oradea | 0.65 | 0.35 |
| Piatra-Neamt | 0.65 | 0.45 |
| Pitesti | 0.6 | 0.5 |
| Ploiesti | 0.6 | 0.35 |
| Ramnicu Valcea | 0.6 | 0.4 |
| Resita | 0.5 | 0.45 |
| Satu Mare | 0.55 | 0.35 |
| Sfantu Gheorghe | 0.55 | 0.35 |
| Sibiu | 0.55 | 0.35 |
| Slatina | 0.55 | 0.45 |
| Slobozia | 0.6 | 0.45 |
| Suceava | 0.55 | 0.4 |
| Targoviste | 0.6 | 0.35 |
| Targu Jiu | 0.55 | 0.4 |
| Targu Mures | 0.5 | 0.35 |
| Timisoara | 0.5 | 0.4 |
| Tulcea | 0.55 | 0.35 |
| Vaslui | 0.5 | 0.45 |
| Zalau | 0.55 | 0.35 |
| Mean | 0.58 | 0.4 |

**Vegetation discrimination threshold vs. non-vegetation – MODIS NDVI**
*Figure 5*

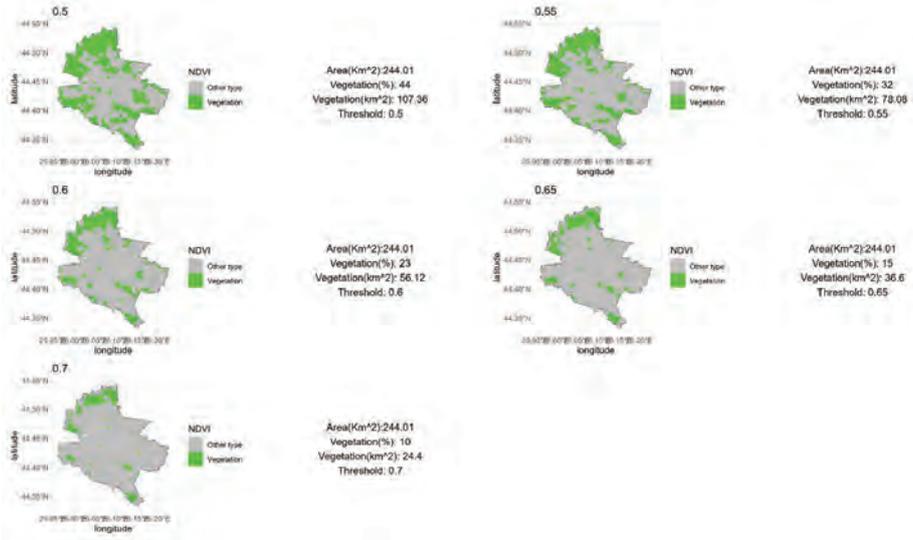

**Histogram of MODIS and Sentinel 2 NDVI values for Bucharest**
*Figure 6*

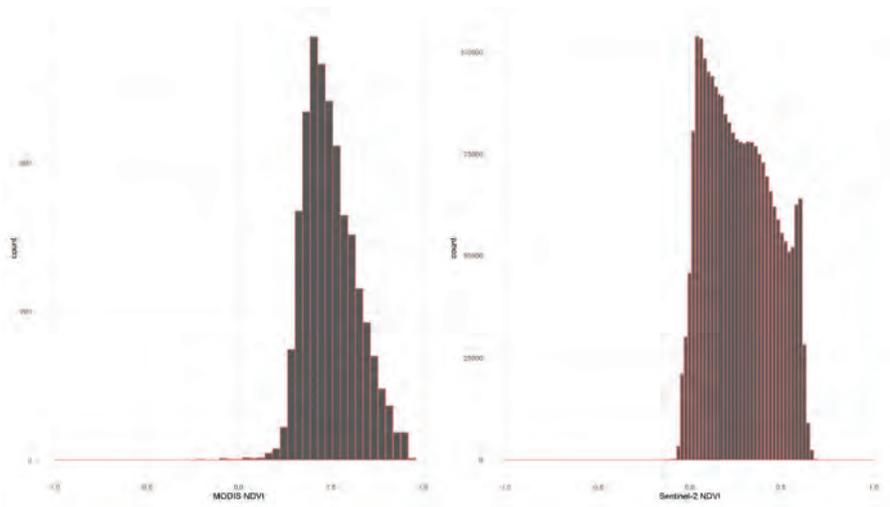

a) Access methods

To access the data we developed two R procedures, one for MODIS and the other for Sentinel 2, using the getSpatialData packages (Schwalb-Willmann, 2022) and the sen2r package (Ranghetti et al, 2022). The download procedure requires a valid account on the EarthData data service, for MODIS, and on the Open Access Hub, for Sentinel 2. The procedure essentially involves selecting a product associated with a mission/sensor (i.e. MODIS TERRA), selecting a window temporal and a region of interest. The native map projection of the data was used for area calculation, sinusoidal projection for MODIS, respectively UTM/WGS84 for Sentinel 2 (it was not necessary to reproject the data for Sentinel 2, given that we did not encounter the problem of a city of residence being located at border region between two adjacent UTM zones).

b) Auxiliary data

As auxiliary data, we used the delimitation of the intra-urban areas of the county-seat cities in vector format, transformed to the native cartographic projection of MODIS and Sentinel data.

c) Data pre-processing.

In the case of MODIS, we extracted the regions of interest from the mosaic raster, applied the discrimination threshold and calculated the area covered by vegetation/the percentage of the total area covered by vegetation. For Sentinel 2 we created a pyramid formed by the spectral bands: B02, B03, B04 and B08, of which we used the first 3 associated with the visible light spectrum to create the "true" color image of the surface of interest, respectively B04 and B08 to calculate the NDVI index, then filtered with the associated discrimination threshold.

The R scripts used for computation are available at the following github address: https://github.com/MarianNecula/NDVI_experimental.git

## 3. RESULTS

Table 3 shows the results obtained after estimating the green index for MODIS and Sentinel 2 data. Between the two sensors, there are differences between the MODIS and Sentinel 2 green index values, either as a result of the spatial resolution, the processed MODIS data having the resolution of 250 meters per pixel, and Sentinel 2 data 10 meters per pixel. For example, in an area of a pixel (square) with a side of 250 meters or a pixel at the MODIS resolution, vegetation represents the majority class (> 50%) covering that area, and the rest <= 50% is represented by other classes, such as built-up areas, then that pixel will be classified as belonging to a vegetated area. In the case of Sentinel 2, the observations being made much more granular, following the same example, the discrimination between a pixel associated with the vegetation class and a pixel associated with another class, is made at the level of a pixel (square) with a side of 10 meters.

**Estimated area of the surfaces covered with vegetation for the county seat cities. -July 2022 Romania**

*Table 3*

| Name | Estimated Surface (km2) | The surface of the vegetation (%) - MODIS | The surface of the vegetation (%) - Sentinel 2 | The surface of the vegetation (km2) - MODIS | The surface of the vegetation (km2) - Sentinel 2 |
|---|---|---|---|---|---|
| Alba Iulia | 15.52 | 23 | 15 | 3.57 | 2.33 |
| Alexandria | 9.62 | 7 | 15 | 0.67 | 1.44 |
| Arad | 45.57 | 13 | 9 | 5.92 | 4.1 |
| Bacău | 34.18 | 27 | 29 | 9.27 | 9.91 |
| Baia Mare | 31.75 | 35 | 31 | 11.12 | 9.85 |
| Bistrița | 16.7 | 31 | 18 | 5.18 | 3.01 |
| Botoșani | 15.37 | 5 | 20 | 0.77 | 3.07 |
| Brăila | 33.17 | 22 | 28 | 7.34 | 9.28 |
| Brașov | 37.55 | 30 | 18 | 11.3 | 6.76 |
| București | 244.51 | 27 | 29 | 66.23 | 70.86 |
| Buzău | 20.52 | 5 | 10 | 1.03 | 2.05 |
| Călărași | 19.41 | 15 | 12 | 2.93 | 2.33 |
| Cluj-Napoca | 93.09 | 59 | 32 | 54.95 | 29.8 |
| Constanța | 44.19 | 4 | 16 | 1.78 | 7.06 |
| Craiova | 44.28 | 12 | 12 | 5.31 | 5.31 |
| Deva | 12.43 | 44 | 36 | 5.47 | 4.47 |
| Drobeta-Turnu Severin | 12.98 | 0 | 6 | 0 | 0.78 |
| Focșani | 11.86 | 2 | 18 | 0.24 | 2.13 |
| Galați | 56.12 | 6 | 18 | 3.39 | 10.09 |
| Giurgiu | 26.39 | 50 | 36 | 13.23 | 9.49 |
| Iași | 46.42 | 29 | 27 | 13.53 | 12.52 |
| Miercurea Ciuc | 9.14 | 45 | 21 | 4.13 | 1.92 |
| Oradea | 54.69 | 28 | 27 | 15.3 | 14.75 |
| Piatra-Neamț | 18.29 | 36 | 28 | 6.61 | 5.12 |
| Pitești | 28.91 | 37 | 25 | 10.72 | 7.23 |
| Ploiești | 50.92 | 40 | 28 | 20.44 | 14.25 |
| Râmnicu Vâlcea | 10.11 | 34 | 26 | 3.44 | 2.63 |
| Reșita | 16.2 | 62 | 40 | 10.03 | 6.47 |
| Satu Mare | 21.81 | 11 | 21 | 2.4 | 4.58 |
| Sfântu Gheorghe | 10.95 | 41 | 26 | 4.5 | 2.84 |
| Sibiu | 25.57 | 8 | 19 | 2.05 | 4.86 |
| Slatina | 17.56 | 9 | 15 | 1.58 | 2.64 |
| Slobozia | 9.68 | 22 | 17 | 2.14 | 1.64 |

| | | | | |
|---|---|---|---|---|
| Suceava | 29.98 | 40 | 31 | 12.03 | 9.29 |
| Târgoviște | 16.06 | 26 | 15 | 4.19 | 2.41 |
| Târgu Jiu | 25.46 | 49 | 20 | 12.48 | 5.09 |
| Târgu Mures | 23.6 | 29 | 16 | 6.85 | 3.78 |
| Timișoara | 68.87 | 23 | 16 | 15.82 | 11 |
| Tulcea | 17.39 | 19 | 13 | 3.33 | 2.26 |
| Vaslui | 9.57 | 29 | 30 | 2.79 | 2.87 |
| Zalău | 17.23 | 30 | 29 | 5.17 | 4.99 |

In Appendix we provide a few maps for the lowest (Drobeta Turnu Severin), highest (Reșița), ranked city, respectively the capital city in terms of vegetation coverage estimates.

**Top county residences according to the percentage of the total area covered with vegetation**

*Figure 4*

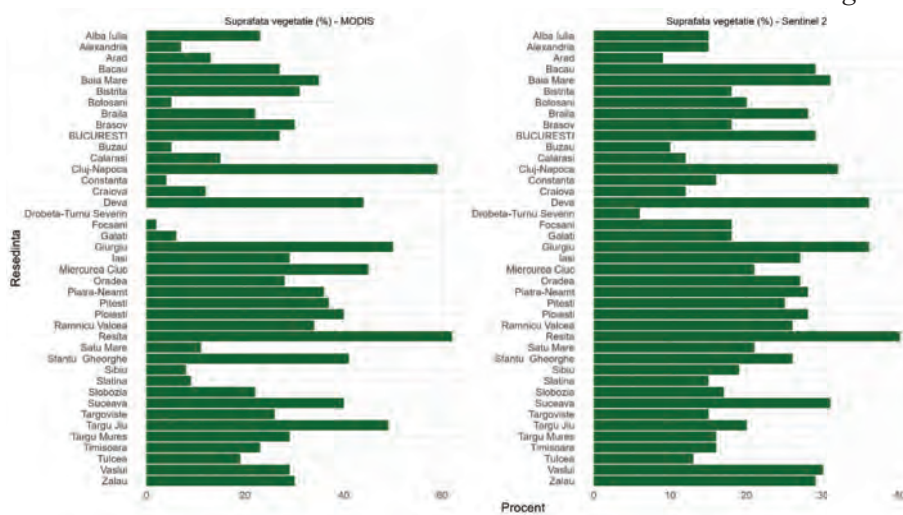

## 4. LIMITATIONS AND DISCUSSIONS

a) The low resolution of MODIS data can be an impediment, when estimates are needed for small areas, where vegetation is interspersed between buildings and is below the area covered by MODIS pixels (<250m). On the other hand, MODIS can provide time series starting from 2000, by comparison with data provided by the Sentinel 2 mission, starting in 2016. From the literature, we identified a series of algorithms that can be used to fuse MODIS data with those of Sentinel 2 to extend the time span while increasing the spatial resolution of the data.

b) Image quality is an important factor. The data can be affected by 2 problems: the complete or partial absence of data from a satellite pass through the area of interest, for example which can be triggered by errors at the sensor level, or the presence of massive clouds that prevent light radiation from reaching to the sensor. The latter type of problem can be partially countered with data coming from another type of sensor, the active monitoring of the earth's surface, of the radar type, which allows the detection of both landforms and natural or anthropogenic elements that cover the earth's surface. In this case, the complexity of data preprocessing increases accordingly, given the need to calibrate the radar signal for a multitude of factors.

c) The construction of a time series for this type of application is considered to identify if there have been substantial changes in the areas covered with vegetation, e.g. the uncontrolled expansion of residential areas within cities as a result of the boom in the real estate market over the last 20 years.

d) There are many other types of applications for green indices in agriculture or forestry that can be developed based on remote sensing data.

e) Within the project we are considering the design of a data dissemination application through a GIS application, with a spatiotemporal selection and comparison functionality between different areas and time periods.

## CONCLUSIONS

This document describes a case study on the use of remote sensing data in official statistics. Starting from a similar project carried out by the Statistics Canada, we used data from the MODIS mission, and additionally data from the Sentinel-2 mission to estimate an urban green index and to make a classification of county residences in Romania according to of the percentage of the city's surface that contains a form of vegetation. At the NIS level (2021) there is a survey that collects and disseminates data on the area covered with vegetation at the level of cities (Verdure spots area in municipalities and towns by macro regions, development regions and counties - matrix GOS103B from the Tempo database), but there are differences at the level of definition of green space from cities. Thus, in the NIS survey, green spaces are considered: parks, public gardens, squares with vegetation, plots with trees and flowers, forests, cemeteries, sports fields and bases, by comparison with the present index that assimilates with vegetation all types of surfaces that contain vegetation, regardless of ownership or use. Using auxiliary data, models or algorithms can be created by which green spaces can be detected or reduced to those listed for the NIS survey to obtain an alternative to the current way of producing this type of statistic. Also, an argument in favor of the exploitation of this data source, the spatial and temporal granularity of the statistics can be considerably increased, from cities and municipalities to areas of arbitrary size, from an annual coverage to a monthly statistic or even with a bi-frequency monthly. A disadvantage is the considerable pre-processing effort, given the size and specificity of these data, while having a set of interdisciplinary knowledge (GIS, remote sensing, statistics). This classification can be used as a test in the experimental statistics section to check the potential interest of statistical data users relative to the advantages/disadvantages of the data source.

**References**

1. European Commission 2014. "Mapping and Assessment of Ecosystems and Their Services." https://ec.europa.eu/environment/nature/knowledge/ecosystem_assessment/pdf/102.pdf.
2. European Commission 2013. "The EU Strategy on Green Infrastructure." https://ec.europa.eu/environment/nature/ecosystems/strategy/index_en.htm.
3. ———. 2016 "The Urban Agenda for EU." https://futurium.ec.europa.eu/en/urban-agenda/pages/what-urban-agenda-eu.
4. ———. 2022 "Copernicus Earth Observation Programme." https://www.copernicus.eu/en/about-copernicus.
5. European Parliament 2013. "7th Environmental Action Programme." https://eur-lex.europa.eu/legal-content/EN/TXT/?uri=CELEX:32013D1386.
6. European Spatial Agency (ESA) 2022a. "2021 Copernicus Sentinel Data Access Annual Report." https://scihub.copernicus.eu/twiki/pub/SciHubWebPortal/AnnualReport2021/COPE-SERCO-RP-22-1312_-_Sentinel_Data_Access_Annual_Report_Y2021_merged_v1.0.pdf
7. ———. 2022b "Green Urban Areas Classification." https://land.copernicus.eu/.
8. ———. 2022c. "Land Monitoring Service." https://land.copernicus.eu/.
9. ———. 2022d. "Open Access Hub" https://scihub.copernicus.eu/dhus/#/home
10. ———. 2022e. "Data Formats" https://sentinels.copernicus.eu/web/sentinel/user-guides/sentinel-2-msi/data-formats
11. ———. 2022f. "Level-2A Algorithm Overview." https://sentinels.copernicus.eu/web/sentinel/technical-guides/sentinel-2-msi/level-2a/algorithm
12. Didan, K. Munoz, A. B. Solano, R. Huete, A. 2015 "MODIS Vegetation Index User's Guide". https://vip.arizona.edu/documents/MODIS/MODIS_VI_UsersGuide_June_2015_C6.pdf
13. HDF Group, 2022. "General HDF5 User's Guide". https://portal.hdfgroup.org/display/HDF5/HDF5+User+Guides
14. Huete, A. Justice, C. Van Leeuwen, W.J.D. 1999 "MODIS vegetation index (MOD13)" https://www.researchgate.net/publication/268745810_MODIS_vegetation_index_MOD13
15. ISO 1989. "Thermal Insulation - Heat Transfer by Radiation - Physical Quantities and Definitions." https://www.iso.org/standard/16943.html.
16. Institutul Național de Statistică, 2014. "Strategia de dezvoltare a Sistemului Statistic Național și a statisticii oficiale a României în perioada 2015-2020". https://insse.ro/cms/files/legislatie/programe%20si%20strategii/Stragtegie_2015-2020.pdf
17. Institutul Național de Statistică, 2021. Suprafața spațiilor verzi în municipii și orașe, pe macroregiuni, regiuni de dezvoltare și județe. http://statistici.insse.ro:8077/tempo-online/#/pages/tables/insse-table

18. National Space Agency (NASA) 2022a. "EarthData" https://search.earthdata.nasa.gov/search
19. National Space Agency (NASA) 2022b. "Moderate Resolution Imaging Spectroradiometer." https://modis.gsfc.nasa.gov/
20. ONU 2022. "Obiectivele Dezvoltării Durabile: Obiectivul 11.7 - Indicatorul 70." https://indicators.report/targets/11-7/.
21. Ranghetti, L. Boschett, M. Nutini, F. Busetto, L. 2022. "sen2r: An R toolbox for automatically downloading and preprocessing Sentinel-2 satellite data" https://sen2r.ranghetti.info/
22. Schwalb-Willmann, J. 2022. "getSpatialData R package". https://github.com/16EAGLE/getSpatialData
23. STATCAN, 2021. "Urban greenness, 2001, 2011 and 2019". https://www150.statcan.gc.ca/n1/pub/16-002-x/2021001/article/00002-eng.htm
24. UNSD Task Team on Earth Observations (2017), "Earth Observations for Official Statistics. Satellite Imagery and Geospatial Data Task Team report". https://unstats.un.org/bigdata/task-teams/earth-observation/UNGWG_Satellite_Task_Team_Report_WhiteCover.pdf
25. Weier, J. Herring, D. 2000 Measuring Vegetation (NDVI & EVI) https://earthobservatory.nasa.gov/features/MeasuringVegetation

**Appendix**

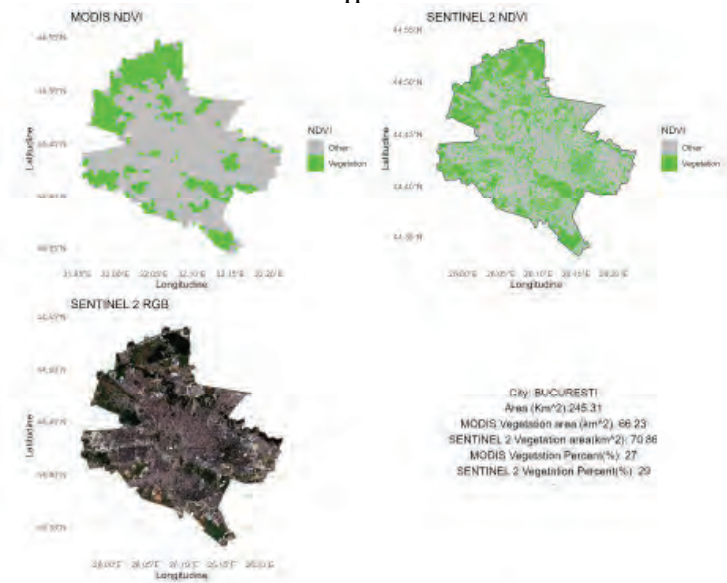

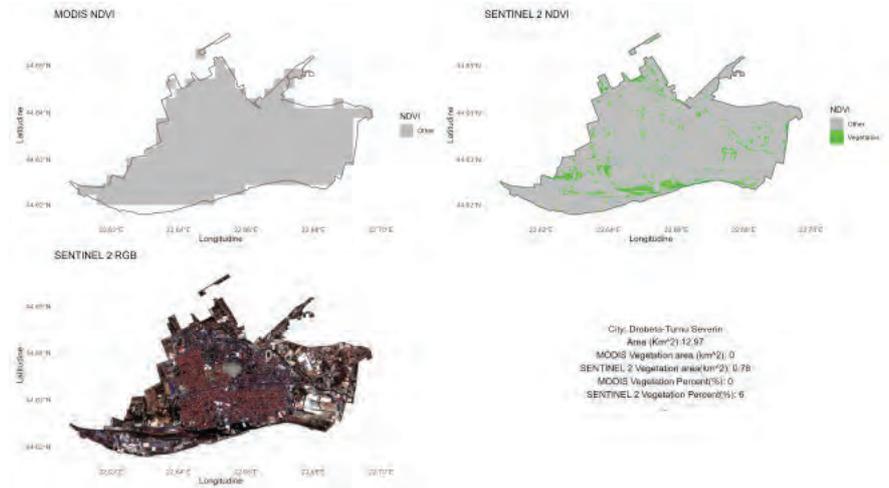
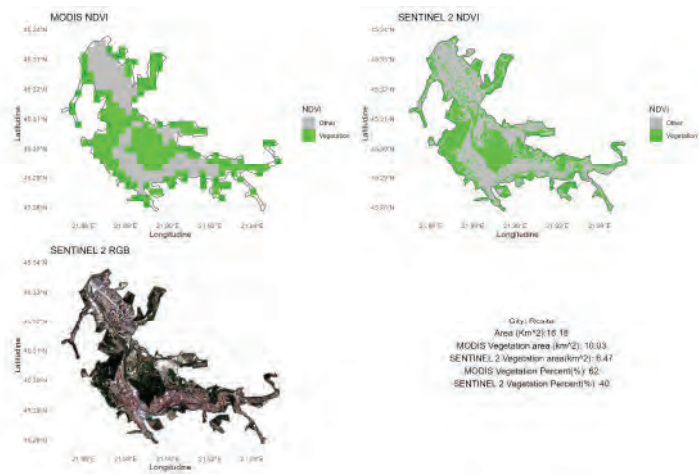